\documentclass[aps,prd,showpacs,superscriptaddress,nofootinbib]{revtex4-2}
\usepackage{graphicx}  
\usepackage{dcolumn}   
\usepackage{bm}        
\usepackage{amssymb}   
\usepackage{amsmath}
\usepackage{xcolor}
\usepackage{rotating}
\usepackage{amsthm}
\usepackage{subfig}
\usepackage{subfloat}
\usepackage{actuarialsymbol}
\usepackage{braket}
\usepackage{mathtools}
\usepackage{bbold}
\usepackage{amssymb}
\usepackage{enumitem}
\usepackage{comment}
\newcommand{\be}{\begin{equation}}\newcommand{\ee}{\end{equation}}
\newcommand{\bea}{\begin{eqnarray}}\newcommand{\eea}{\end{eqnarray}}
\newcommand{\brr}{\begin{array}}\newcommand{\err}{\end{array}}
\newcommand{\bit}{\begin{itemize}}\newcommand{\eit}{\end{itemize}}
\newcommand{\ben}{\begin{enumerate}}\newcommand{\een}{\end{enumerate}}

\newcommand{\ba}{\begin{array}}
\newcommand{\ea}{\end{array}}

\definecolor{darkred}{rgb}{.8,0,0}

\definecolor{darkblue}{rgb}{0,0,.7}

\def\1{{_{1}}}\def\2{{_{2}}}

\def\noHe0{:\;\!\!\;\!\!:H_e(0):\;\!\!\;\!\!:}
\def\noHm0{:\;\!\!\;\!\!:H_\mu(0):\;\!\!\;\!\!:}

\def\1{{_{1}}}\def\2{{_{2}}}

\begin{document}
\title{Entanglement saturation in quantum electrodynamics scattering processes}

\author{Massimo Blasone}
\email{blasone@sa.infn.it}
\affiliation{Dipartimento di Fisica, Universit\`a di Salerno, Via Giovanni Paolo II, 132 I-84084 Fisciano (SA), Italy}
\affiliation{INFN, Sezione di Napoli, Gruppo collegato di Salerno, Italy}

\author{Silvio De Siena}
\email{silvio.desiena@gmail.com}
\affiliation{(Ret. Prof.) Universit\`a di Salerno, Via Giovanni Paolo II, 132 I-84084 Fisciano (SA), Italy}

\author{Gaetano Lambiase}
\email{lambiase@sa.infn.it}
\affiliation{Dipartimento di Fisica, Universit\`a di Salerno, Via Giovanni Paolo II, 132 I-84084 Fisciano (SA), Italy}
\affiliation{INFN, Sezione di Napoli, Gruppo collegato di Salerno, Italy}

\author{Cristina Matrella}
\email{cmatrella@unisa.it}
\affiliation{Dipartimento di Fisica, Universit\`a di Salerno, Via Giovanni Paolo II, 132 I-84084 Fisciano (SA), Italy}
\affiliation{INFN, Sezione di Napoli, Gruppo collegato di Salerno, Italy}

\author{Bruno Micciola}
\email{bmicciola@unisa.it}
\affiliation{Dipartimento di Fisica, Universit\`a di Salerno, Via Giovanni Paolo II, 132 I-84084 Fisciano (SA), Italy}
\affiliation{INFN, Sezione di Napoli, Gruppo collegato di Salerno, Italy}
\affiliation{Fakult\"{a}t f\"{u}r Mathematik, Universit\"{a}t Wien, Oskar-Morgenstern-Platz 1, 1090 Vienna, Austria}

\date{\today}

\vspace{1cm}

\begin{abstract}
We investigate the properties
of quantum electrodynamics (QED) two-particle scattering processes
when an arbitrarily
sharp filtering of the outgoing particles in momentum space
is performed. 
We find that  these processes are described 
by dynamical quantum maps, whose structure is such that any initial state is transformed into a maximally entangled state, after an infinite number of iterations of the map.
This structural property
is exactly realized if all the colliding particles
are massive fermions
while, when photons are involved, it is verified in a partial way, depending on the process under consideration.
\end{abstract}

\maketitle

\section{Introduction}

Since the dawn of quantum mechanics, scattering processes of elementary particles
have been at the heart of modern physics
as the main tool to investigate
fundamental phenomena at different energy scales. 
The recent development of quantum information promotes  powerful
technological advancements, parallel to an ever deeper conceptual understanding of  foundational properties of quantum mechanics \cite{NeCQI}.
Two topics whose roots are placed in the more hidden facets of quantum phenomena,
were inevitably bound to intersect, as e. g. it happened in the Higgs boson detection
at CERN, in which the concept of entanglement played an important role in
decoding data\footnote{``In this experiment, correlations arising from quantum entanglement 
of the four charged leptons generated after weak decaying, 
were used in the CMS discovery analysis 
to help distinguish the Higgs signal from backgrounds; 
had we not so used our understanding of entanglement, 
there would not have been a CMS discovery announcement on July 4, 2012"
\cite{Lykken}.}.
The idea has thus developed that, exploiting the methods of quantum information
to investigate quantum correlations
in scattering processes, can shed new light on the
deepest inner structure of these processes,
and on possible effects beyond the Standard Model \cite{Shi:2004yt}-\cite{Carena:2025wyh}.
Since nonlocal correlations have taken on a crucial role
in highlighting the clearest breaking between classical and quantum world,
much attention has been focused on entanglement generation and modification
following scattering processes. So,
increasingly systematic studies have been devoted to this topic
for different fundamental interactions\cite{Kharzeev,Cervera1,Beane,Araujo,Fan,Fan:2017mth,Afik:2022kwm,Fonseca:2021uhd,Afik:2023dgh,Sinha:2022crx,Serafini,Blasone:2024dud,BlasoneSE2}, and a first experimental verification
has been obtained for quarks \cite{Atlas}.

In Refs.\cite{Cervera1,Serafini,Blasone:2024dud,BlasoneSE2} 
entanglement in QED processes has been investigated 
by focusing on scattering at fixed momentum, requiring in center of mass (COM) reference frame to specify only
a scattering angle and the modulus of the momentum\footnote{Actually, one fixes the ratio $\mu =|p|/m$,
with $m$ the mass of incoming particles, a parameter that rules the transition
from the non relativistic regime (low values of $\mu$) to the ultrarelativistic one
("$\mu \rightarrow \infty$").}. Actually, investigations in the COM of scattering angle 
regions is what is typically performed in scattering experiments. In the above references, thus, 
it is assumed that an arbitrarily
sharp filtering of the outgoing particles in momentum space
is performed, without resolving their internal (helicity or
polarization) degrees of freedom. This corresponds
to a positive operator valued measurement (POVM) leading to
a post-measurement helicity state \cite{Serafini}.

In the framework of this approach, it has been recently shown \cite{BlasoneSE2} that,
at tree level, QED scattering processes completely preserve maximum entanglement,
as long  as colliding particles associated with different statistics 
are not involved\footnote{Maximum entanglement is
totally conserved when all the input and output particles are massive fermions, 
while, when photons are involved, 
maximum entanglement is only partially preserved, as in the annihilation
in two photons, or not preserved at all, as in the Compton scattering in which 
fermions and photons interact directly \cite{BlasoneSE2}.}. This result brings
back to the question of the quantum entropy gain under the action of quantum maps, which has
been investigated in literature, in particular in Refs.\cite{Holevo,Lesovik}: A lower bound on the quantum entropy gain under the
action of a general quantum channel (a completely positive, trace preserving (CPTP) map) can be further improved for a unital quantum channel to provide a zero lower
bound, leading to a quantum version of the H-Theorem. For unital quantum
channels, in particular for unitary evolutions as those described by the S-matrix, maximally entangled states are thus
fixed points. However, the POVM realizing an arbitrarily sharp filtering of the outgoing particles in momentum
space, introduces a substantial modification because it does not implement a unital quantum channel \footnote{A \emph{quantum channel} is defined 
as a trace-preserving completely positive map (CSTP) $\Phi (\rho)$ 
of a density matrix $\rho$. A \emph{unital quantum channel} is a quantum channel 
that satisfies $\Phi (\textbf{1}) = \textbf{1}$. Obviously, 
an unitary operator is a particular case of unital quantum channel: $U^{- 1} \textbf{1} U = \textbf{1}$. 
The lower bound on the entropy gain is: $S \Phi [(\rho)] – S (\rho) \ge – k_B Tr \{\Phi (\rho) \log{\Phi (\textbf{1})\}}$, 
where $S$ is the entropy. For a unital quantum channel $\Phi (\textbf{1}) = 0$, 
and the lower bound becomes zero providing a quantum version of the $H$-theorem.}, and the above lower bounds do not apply. In fact, as shown in Ref.\cite{BlasoneSE2}, entanglement can
also decrease after scattering. Despite of this, in Ref.\cite{BlasoneSE2} and in more complete
form in Ref.\cite{NoiChaos}, we proved that after the POVM, maximal entanglement is conserved, exclusively as a consequence
of the general form of the dynamical quantum maps which describe
the effects of the scattering processes. In this paper we show that, for 2 fermions $\rightarrow$ 2 fermions scattering, repeated iterations of the above quantum maps (performed both at fixed angle and by randomly varying the angle at each step) leads to a saturation of the entanglement, while this result can be reduced, or deleted, if also photons are involved. We also analyze the underlying mechanism for which maximal entanglement is obtained by repeated applications of the maps.

\section{Formalism and assumptions}

Here we provide only a glimpse on the formalism. 
Scattering processes can be accounted for by the $S$-matrix formalism,
which describes the unitary evolution of the system from an initial state
$\rho_i$ at $t = - \infty$
to the final state $\rho_f$  at $t = \infty$, after scattering has occurred.
In the COM reference frame
of the colliding particles, we study the spin degrees of freedom between states
before and after the scattering process at fixed momenta, 
i.e.  it is assumed that an arbitrarily
sharp filtering of the outgoing particles in momentum space
is performed, without resolving their internal (helicity or
polarization) degrees of freedom.
This corresponds
to perform a POVM on the final state $\rho_f$ after the unitary evolution, leading to
a post-measurement helicity state. After performing this procedure, one can exploit 
perturbative approximations and investigate  nonlocal helicity correlations
for specific QED scattering processes (see for example Section II.A of Ref. \cite{Serafini}
for a more complete description).

Assuming an initial pure state, the final post-measurement state, after normalization, is again a pure state,
which can be described by a density matrix 
or by a vector state.
For a generic initial state, the final states can be expressed in terms 
of the scattering amplitudes $\mathcal{M}_{a,b;r,s}$,
with $a, b$ and $r, s$ denoting the initial and final helicities, respectively.
Scattering amplitudes are functions of the scattering angle $\theta$, and of the modulus of the momentum $|p|$
through the parameter $\mu =|p|/m$, and are here computed at tree level.

\section{Quantum maps and entanglement saturation}

In Ref. \cite{NoiChaos} QED scattering processes after momentum filtering have been described
as dynamical quantum maps. Describing input and output states as vectors 
in the Hilbert space, these maps can be defined by the scattering-amplitude matrices
(displaying a different form for each process)
\be
\textbf{M} = \left(
\begin{array}{cccc}
\mathcal{M}_{RR; RR} & \mathcal{M}_{RL; RR} & \mathcal{M}_{LR; RR} & \mathcal{M}_{LL; RR} \\
\mathcal{M}_{RR; RL} & \mathcal{M}_{RL; RL} & \mathcal{M}_{LR; RL} & \mathcal{M}_{LL; RL} \\
\mathcal{M}_{RR; LR} & \mathcal{M}_{RL; LR} & \mathcal{M}_{LR; LR} & \mathcal{M}_{LL; LR} \\
\mathcal{M}_{RR; LL} & \mathcal{M}_{RL; LL} & \mathcal{M}_{LR; LL} & \mathcal{M}_{LL; LL}
\end{array}
\right).
\label{SAM}
\ee
The maps must be applied to an initial input
state to provide, after normalization depending on the initial state, the final state after scattering and POVM procedure.
In Ref. \cite{NoiChaos} the invariant sets for the above maps have been identified. Invariant sets
include the entire set of maximally entangled states
if both input and output particles
are fermions, and a subset of maximally entangled states
in the case of scattering $e^{-} e^{+} \rightarrow \gamma \gamma$,
while in the case of Compton scattering, maximal  entangled states is not preserved.
These results are obtained  by resorting only to  the shape of the maps, 
which is a consequence of the relations among scattering amplitudes, without reference to their
explicit expressions.

A notable result can be obtained
iterating the maps.
Starting by some initial state $\ket{i}$, we can consider the iterative map
\begin{equation}
    \ket{f_{n + 1}} = \textbf{M} \ket{f_n},
\end{equation}
which, if $\ket{f_0} = \ket{i}$,  leads to $\ket{f_n} = \textbf{M}^n \ket{i}$, 
providing a ``trajectory" in Hilbert space. In implementing this procedure, it is easily verifiable
that introducing normalization step by step is equivalent to directly normalizing the state after the $n$-th iteration.
The map can be iterated both by keeping the angle fixed at each step, and by randomly changing the angle after each step.
We show that, due to the properties of the scattering matrices, entanglement is saturated,
apart in the few cases in which there is no entropy gain at the first step.

\begin{figure}[t]
    \centering
    \includegraphics[width = 10.0 cm]{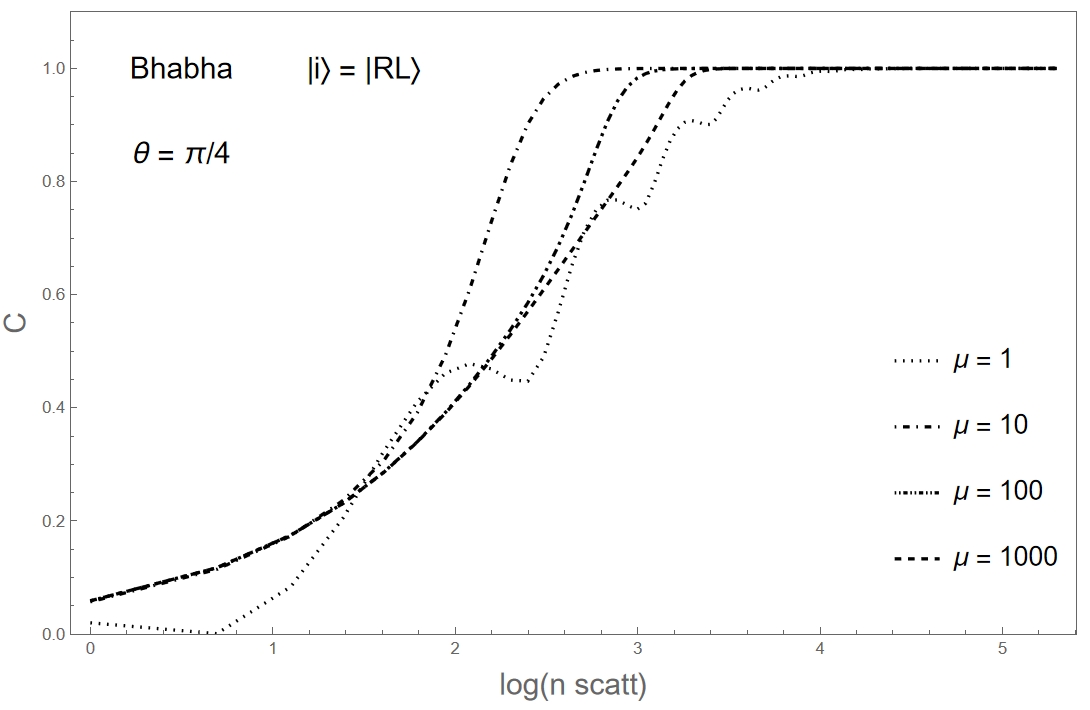}
    \caption{Concurrence in iterated Bhabha process for different incoming momenta with $\ket{RL}$ as initial state and scattering angle $\theta = \pi/4$.}
    \label{fig1}
\end{figure}

\begin{figure}[t]
    \centering
    \includegraphics[width = 10.0 cm]{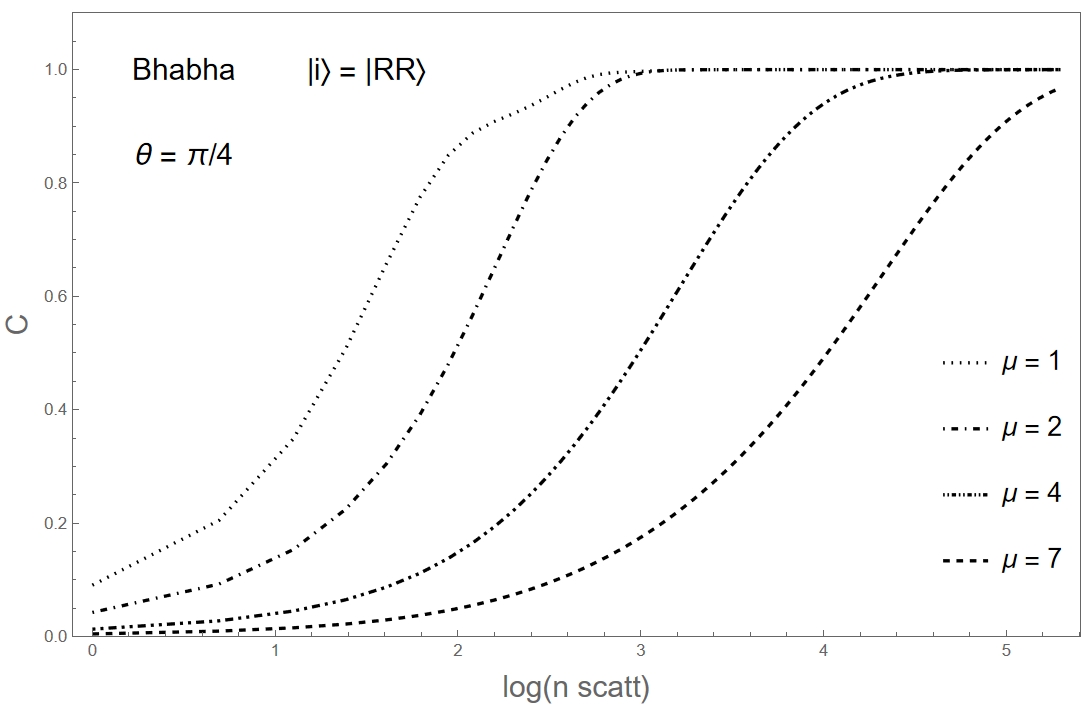}
    \caption{Concurrence in iterated Bhabha process for different incoming momenta with $\ket{RR}$ as initial state and scattering angle $\theta = \pi/4$.}
    \label{fig2}
\end{figure}

\begin{figure}[h]
    \centering
    \includegraphics[width = 10.0 cm]{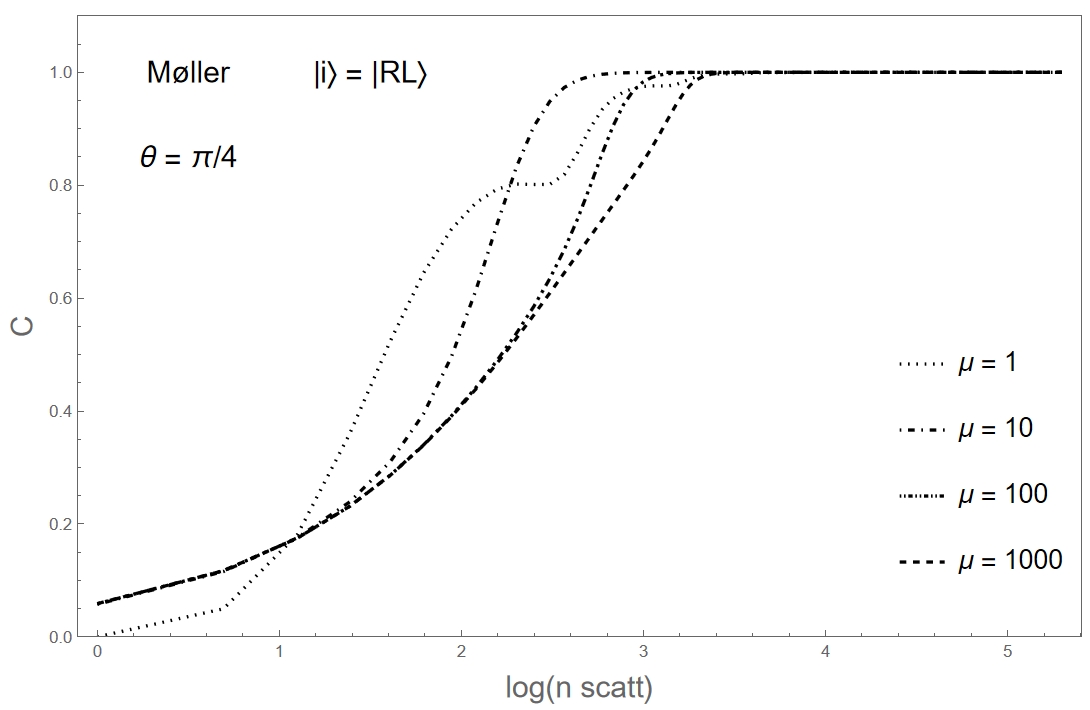}
    \caption{Concurrence in iterated M\o{}ller process for different incoming momenta with $\ket{RL}$ as initial state and scattering angle $\theta = \pi/4$.}
    \label{fig3}
\end{figure}

\begin{figure}[h]
    \centering
    \includegraphics[width = 10.0 cm]{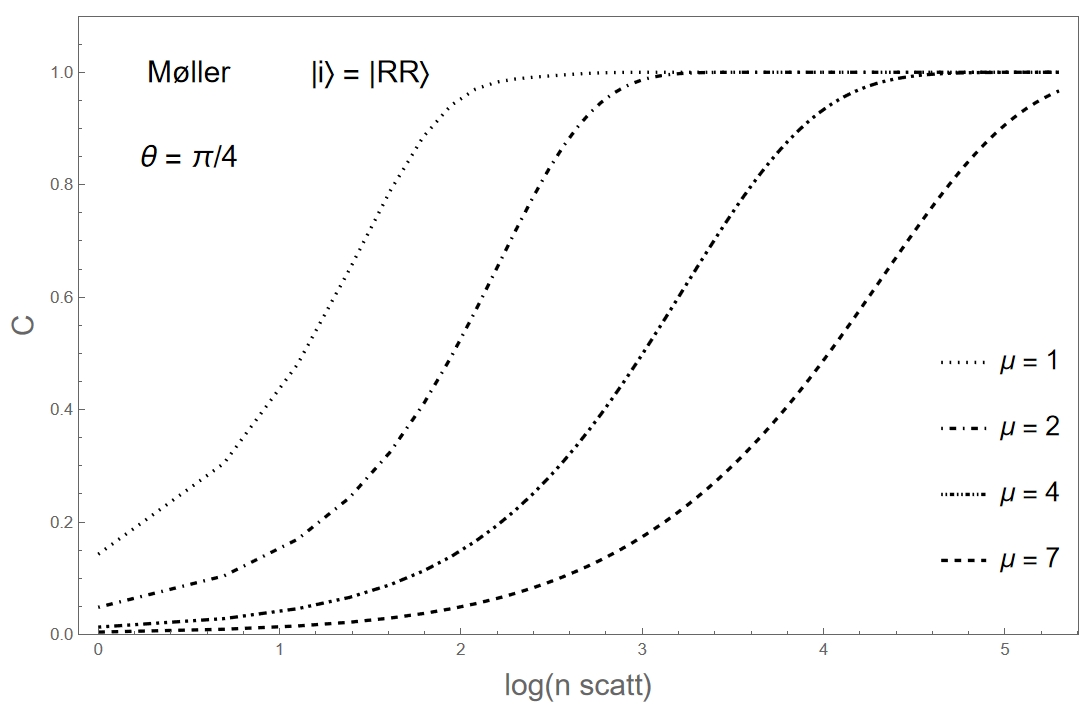}
    \caption{Concurrence in iterated M\o{}ller process for different incoming momenta with $\ket{RR}$ as initial state and scattering angle $\theta = \pi/4$.}
    \label{fig4}
\end{figure}

\begin{figure}[h]
    \centering
    \includegraphics[width = 10.0 cm]{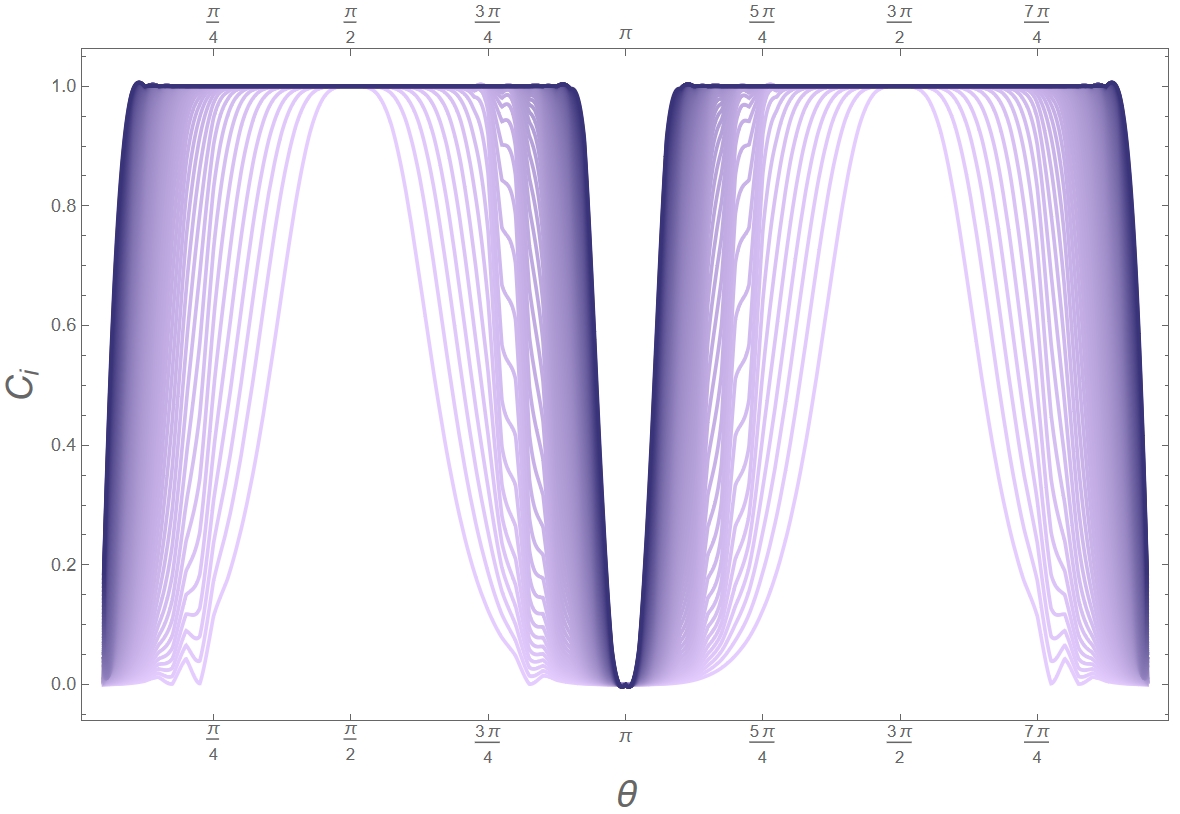}
    \caption{Concurrence in iterated Bhabha process for $\mu=10$ and $\ket{RL}$ as initial state for all the scattering angles.}
    \label{fig5}
\end{figure}

\begin{figure}[h]
    \centering
    \includegraphics[width = 10.0 cm]{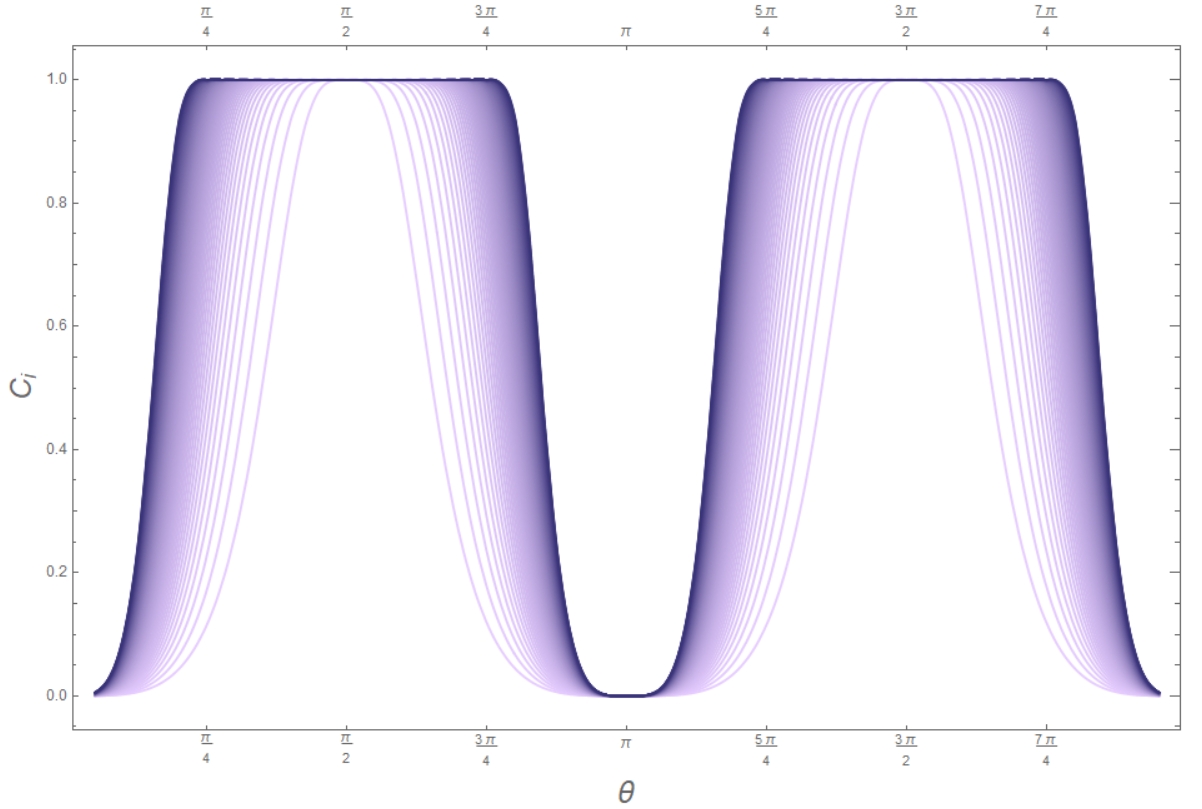}
    \caption{Concurrence in iterated Bhabha process in the ultra-relativistic regime with $\ket{RL}$ as initial state for all the scattering angles.}
    \label{fig6}
\end{figure}

We first fix the scattering angle at each step: After the first step, we select the output state
generated at a specific angle, which becomes the input state at the same angle in the second interaction, 
and so on\footnote{We assume a conceptual point of view, regardless of the 
practical difficulty to realize iterations. In fact, we want to highlight structure of
all maps that encodes the asymptotic increase of entanglement.}. We provide 
numerical evidence when the simplest factorized initial states
are assumed.
In Figures \ref{fig1}, \ref{fig2}, \ref{fig3} and \ref{fig4}  we can see
the asymptotic trend to saturation of the concurrence for different values of $\mu$ if one iterates the maps associated to the Bhabha and M\o{}ller scattering,
starting from $\ket{RL}$  ($\ket{RR})$: 
$\lim_{n \rightarrow \infty} N_{n}^{-1} \, \textbf{M}^n \, \ket{RL} (\ket{RR})$, in which
$N_n$ represents the normalization factor at step $n$. 
In the non relativistic regime, we observe a  non-monotonic behavior due to the contribution of states
associated to a decrease of the entanglement (\emph{entanglophobous} states) \cite{BlasoneSE2},
but anyway the envelops converge to maximum entanglement. In Figs. \ref{fig2} and \ref{fig4} we consider as initial state $\ket{i}=\ket{RR}$, and the entanglement decreases for growing values of $\mu$ until it vanishes; in fact, for such an initial state, in the ultrarelativistic limit $M(RR, rs) = M(rs, RR) = 0, \; \forall r, s \neq RR$, i.e. only the amplitude $M(RR,RR)$ that maps the state onto itself survives.
In Figs. \ref{fig5} and \ref{fig6}, for $\mu=10$ and $\mu=\infty$ with $\ket{i}=\ket{RL}$, we show that the entanglement saturation is reached for all the scattering angles. In the case of the ultra-relativistic limit, for $\theta = \pi$, entanglement is zero because no entanglement is generated after the first step.
\begin{figure}[h]
    \centering
    \includegraphics[width = 10.0 cm]{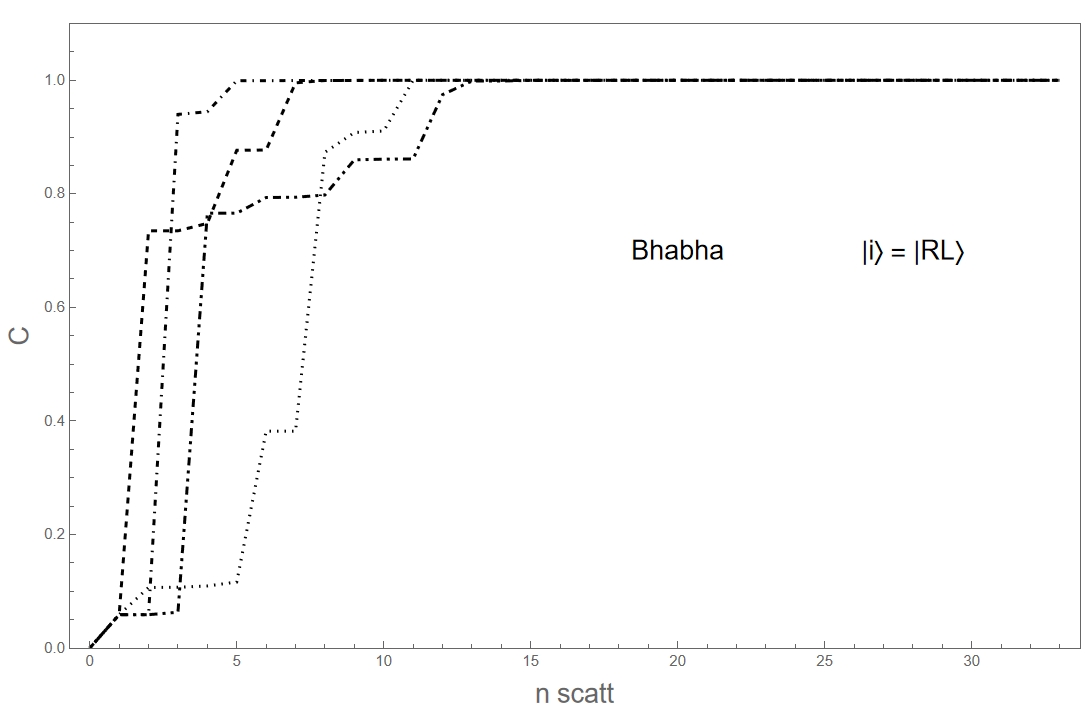}
    \caption{Concurrence in iterated Bhabha process in the ultra-relativistic regime with $\ket{RL}$ as initial state for each scattering angle.}
    \label{fig7}
\end{figure}

In Table \ref{tab1} we report the asymptotic states
obtained in both regimes
for Bhabha and M$\o{}$ller scattering processes, 
if the initial states are $\ket{RL}$ and $\ket{RR}$.
We see that such states are the maximally entangled states included in the Tables of Ref.\cite{BlasoneSE2}. 
Bell states are asymptotic states independent of the scattering angle,
while in the superpositions of Bell states the angles $s_i$ 
change their values when the scattering angle $\theta$ changes.

\begin{table}[t]
    \begin{tabular}{|l|l|l|r|}
\hline
SCATTERING PROCESS & INITIAL STATE(S) & REGIME & ASYMPTOTIC STATE \\
Bhabha & $\ket{RL}$ & u. r. & $\Psi^{+}$ \\
\hline
\hline
Bhabha & $\ket{RL}$ & n. r. & $\cos{s_1}\Phi^{-}+\sin{s_1}\Psi^+$ \\
\hline
\hline
Bhabha & $\ket{RR}$ & n. r. & $\Phi^{+}$ \\
\hline
\hline
M\o{}ller & $\ket{RL}$ & u. r. & $\Psi^{-}$ \\
\hline
\hline
M\o{}ller & $\ket{RR}$ & n. r. & $\Phi^{-}$ \\
\hline
\hline
M\o{}ller & $\ket{RL}$ & n. r. & $\cos{s_3} \Phi^{+} + \sin{s_3} \Psi^{-}$ \\ 
\hline
\end{tabular}
    \caption{Some asymptotic, maximally entangled states after infinite iteration of the dynamical quantum maps 
    for fermions-fermions scattering processes,
    where $\Phi^{\pm}, \Psi^{\pm}$ are the Bell states.
    Here u. r. denotes the ultrarelativistic regime, and n. r. the nonrelativistic one. }
    \label{tab1}
\end{table}

We can also iterate the map describing the Bhabha scattering
by starting from an initial angle
and then randomly changing the angle at each
step.
In Fig. \ref{fig6} we plot, by
repeating the random procedure, the concurrence obtained for $\ket{i}=\ket{RL}$ 
in the case of Bhabha scattering in the ultra-relativistic regime.
We see that also in this case, the entanglement is asymptotically saturated.

\section{Entanglement saturation mechanism}

We can now go into more detail about the game. 
As an exemplary case, we take the Bhabha scattering process. Its action can be summed up by the
dynamical quantum map defined by the matrix \cite{NoiChaos}
\be
\textbf{M} = \left(
\begin{array}{cccc}
A & - B & - B & D \\
B & E & F & - B \\
B & F & E & - B \\
D & B & B & A
\end{array}
\right).
\label{MB}
\ee
where $A = \mathcal{M}_{RR; RR}, \; B = \mathcal{M}_{RR; RL}, \; D = \mathcal{M}_{RR; LL}, \;
E = \mathcal{M}_{RL; RL}, \; F = \mathcal{M}_{RL; LR}$.
Now, we highlight three relevant properties of all the quantum maps
epitomizing the scattering processes of 2 fermions $\rightarrow$ 2 fermions, and which
hold only by their general form, regardless of the explicit dependence of their elements
(scattering amplitudes)
on the regime or on the scattering angle:

\begin{enumerate}
    \item 
The set of the maximally entangled states is an invariant set for the maps defined by the
scattering matrices $\textbf{M}$.

\item
The powers $\textbf{M}^n$ hold the same form of the original matrices for any value of $n$
(self-similarity by raising to a generic power). 
For example, in the case of the Bhabha process we have
\be
\textbf{M}^n = \left(
\begin{array}{cccc}
A_n & - B_n & - B_n & D_n \\
B_n & E_n & F_n & - B_n \\
B_n & F_n & E_n & - B_n \\
D_n & B_n & B_n & A_n
\end{array}
\right),
\label{MBn}
\ee
where any element in this matrix is expressed\footnote{We note that, due to this property, the states obtained 
applying $n$ times the matrix on the factorized initial states $\ket{RR}, \ket{RL}, \ket{LR}, \ket{LL}$,
are given, before normalization, by the column vectors defined by the first to the fourth 
column of matrix (\ref{MBn}), respectively.} in terms of the original elements of
the matrix Eq.(\ref{MB}).

\item 
The infinite iteration of the map on an initial state converges to a maximally entangled state,
ensuring entanglement saturation.

\end{enumerate}

The first property, as already pointed out, has been proved in Ref. \cite{NoiChaos}, and
the second property can be easily verified. 
We now show how the third property is realized for the case of Bhabha scattering
(for the other 2 fermions $\rightarrow$ 2 fermions scattering 
processes the procedure is similar).

The starting point is the fact that the eigenvectors of the scattering matrices
are linearly independent vectors and thus
any state can be expressed as their linear combination.
In the ultrarelativistic regime the procedure is highly simplified,
because the matrices become Hermitian, and define an orthonormal basis
made up of four maximally entangled states (we will provide an example
later). For the non relativistic regime, the situation is more complicated. The Bhabha scattering matrix (\ref{MB}) has four eigenvectors
$\ket{\lambda_1}, \ket{\lambda_2}, \ket{\lambda_3}, \ket{\lambda_4}$, associated to the
eigenvalues $\lambda_1, ..., \lambda_4$.
The first two eigenvectors are: the Bell state $\ket{\lambda_1} = \Phi^{+}$, with eigenvalue
$\lambda_1 = A + D$, and the Bell state $\ket{\lambda_2} = \Psi^{-}$, with eigenvalue
$\lambda_1 = E - F$. The other two eigenvectors computed before normalization, 
and which we denote with $\ket{\lambda_3}', \ket{\lambda_4}'$, 
can be written in the following form
\be
\ket{\lambda_3}' = \left(
\begin{array}{c}
- 1 \\
s_1 + \sqrt{t} \\
s_1 + \sqrt{t} \\
1
\end{array}
\right), \qquad
\ket{\lambda_4}' = \left(
\begin{array}{c}
- 1 \\
s_1 - \sqrt{t} \\
s_1 - \sqrt{t} \\
1
\end{array}
\right)
\label{NMEE}
\ee
where $s_1= A - D + E + F; \; \; t = (- A + D + E + F)^2 - 16 B^2$.
The corresponding eigenvalues are
\be
\lambda_3 = s_1 + \sqrt{t}, \qquad \lambda_4 = s_1 - \sqrt{t}.
\label{EV34}
\ee
Since we want to prove the third property above without resorting to
the explicit expressions of the elements $A, ... , F$, 
we must distinguish two cases; $t \ge 0$ or $t < 0$.
In both cases, we prove our thesis.

\vspace{0.3cm}

If $t \ge 0$, the eigenvalues are real valued, as well as all the coefficients in
vectors (\ref{NMEE}) and, after normalization, 
the four states are:
\begin{equation}
    \ket{\lambda_1} = \Phi^{+}, \quad \ket{\lambda_2} = \Psi^{-},  \quad  
\ket{\lambda_3} = \cos{\delta_3} \Phi^{-} + \sin{\delta_3} \Psi^{+},  \quad  
\ket{\lambda_4} = \cos{\delta_4} \Phi^{-} + \sin{\delta_4} \Psi^{+},
\end{equation}
with angles $\delta_3, \delta_4$ which can be expressed in terms of the elements
$A, ... , F$. All four states are maximally entangled states \cite{BlasoneSE2}.
Any initial state can be written as linear combination
of these linearly independent states. For example, consider the separable
initial state $\ket{i} = \cos{\alpha} \ket{RR} + \sin{\alpha} \ket{RL}$.
We can write
\be
\ket{RR} = \frac{1}{\sqrt{2}} (\Phi^{+} + \Phi^{-}) =
\frac{1}{\sqrt{2}} [\Phi^{+} +  c_3 \ket{\lambda_3} + c_4 \ket{\lambda_4}],
\label{BRR}
\ee
and
\be
\ket{RL} = \frac{1}{\sqrt{2}} (\Psi^{+} + \Psi^{-}) =
\frac{1}{\sqrt{2}} [\Psi^{-} + c_{3}' \ket{\lambda_3} + c_{4}' \ket{\lambda_4}],
\label{BRL}
\ee
where the coefficients $c_3, c_4, c_{3}', c_{4}'$ satisfy
\bea
&&c_{3} \sin{\delta_3} + c_{4} \sin{\delta_4} = 0; 
\quad c_{3} \cos{\delta_3} + c_{4} \cos{\delta_4} = 1;\nonumber \\ 
&&c_{3}' \cos{\delta_3} + c_{4}^{'} \cos{\delta_4} = 0; 
\quad c_{3}' \sin{\delta_3} + c_{4}^{'} \sin{\delta_4} = 1.
\label{ci}
\eea

If we iterate the action of $\textbf{M}$ by keeping fixed the scattering angle at each step, we have
\be
\mathcal{N}_{n}^{- 1} \textbf{M}^n \ket{RL} = 
\frac{1}{\sqrt{2}} \mathcal{N}_{n}^{- 1} [\cos{\alpha} \, \lambda_{1}^{n} \Phi^{+} +
\sin{\alpha} \, \lambda_{2}^{n} \Psi^{-} 
+ (\cos{\alpha} \, c_3 + \sin{\alpha} \, c_{3}') \lambda_{3}^{n} \ket{\lambda_3} 
+ (\cos{\alpha} \, c_4 + \sin{\alpha} \, c_{4}') \lambda_{4}^{n} \ket{\lambda_4}]
\label{ASBRL}
\ee
with $\mathcal{N}_n$ the normalization factor.
In the limit $n\rightarrow \infty$, the eigenvector with dominant eigenvalue will survive and will determine the asymptotic state,
which is anyway a maximally entangled state.

\vspace{0.3cm}

If $t < 0$, we can write
\be
\ket{\lambda_3}' = \left(
\begin{array}{c}
- 1 \\
s_1 + i s_2 \\
s_1 + i s_2 \\
1
\end{array}
\right), \qquad
\ket{\lambda_4}' = \left(
\begin{array}{c}
- 1 \\
s_1 - i s_2 \\
s_1 - \i s_2 \\
1
\end{array}
\right)
\label{NMEE1}
\ee
with $s_2 = \sqrt{|t|}$.
The corresponding eigenvalues are
\be
\lambda_3 = s_1 + i s_2, \qquad \lambda_4 = s_1 - i s_2 \equiv \lambda_{3}^{*},
\label{EV34a}
\ee
which we can also write as
\be
\lambda_3 = r e^{i \eta}, \qquad \lambda_4 = r e^{- i \eta}, 
\qquad r = \sqrt{s_{1}^{2} + s_{2}^{2}}, \qquad \eta = \arctan{\frac{s_2}{s_1}}.
\label{EV34}
\ee
In this case, the eigenvalues are complex valued and the two eigenvectors are not maximally entangled states.
Therefore, we define two new (normalized) states
\begin{eqnarray}
    \Xi_{3} &\equiv & \frac{1}{2}(2 + s_{1}^{2})^{- \frac{1}{2}} 
\big[\ket{\lambda_{3}'} + \ket{\lambda_{4}'}\big] = \cos{\beta} \, \Phi^{-} + \sin{\beta} \, \Psi^{+},
\nonumber \\
\Xi_{4} &\equiv & - i \,(2 s_2)^{- 1} \big[\ket{\lambda_{3}'} - \ket{\lambda_{4}'}\big] = \Psi^{+},
\label{NLIS}
\end{eqnarray}
where we can write the last expression for the state $\ket{\Xi_{3}}$, with a suitable angle $\beta$
expressible as usual in terms of the matrix elements,
because it is a linear combination
of $\Phi^{-}$ and $\Psi^{+}$ with real coefficients
(note that $\cos{\beta}$ cannot be zero).
Now we express an initial state in terms of the four states:
\begin{equation}
    \Phi^{+}, \qquad \Psi^{-}, \qquad \Xi_{3}, \qquad \Xi_{4}.
\end{equation}
They are all maximally entangled states.
Consider again the separable initial state $\ket{i} = \cos{\alpha} \ket{RR} + \sin{\alpha} \ket{RL}$.
We have
\be
\ket{i} = \frac{1}{\sqrt{2}} \Big[\cos{\alpha} \, \Phi^{+} + \sin{\alpha} \, \Psi^{-} 
+ \cos{\alpha} \, (\cos{\beta})^{- 1} \, \Xi_{3} + \sin{\alpha} \, (\cos{\beta})^{- 1} \, \Xi_{4}\Big].
\label{BRRRLN}
\ee
Application of $\textbf{M}^n$ to $\Xi_3$ and $\Xi_4$ gives
\bea
&&\textbf{M}^n \, \Xi_{3} = 
(2 + s_{1}^{2})^{- \frac{1}{2}} \, r^n \, [\cos{(n \eta)} \, \Phi^{-} + (s_1 \cos{(n \eta)} - s_2 \sin{(n \eta)}) \, \Psi^{+}] \nonumber \\
&&\textbf{M}^n \, \Xi_{4} = (s_2)^{- 1} \, r^{n} \, [\sin{(n \eta)} \, \Phi^{-} + (s_1 \sin{(n \eta)} + s_2 \, \cos{(n \eta)}) \, \Psi^{+}]
\label{MnXi}
\eea
where we have exploited expressions (\ref{EV34}) for $\lambda_3, \lambda_4$.
Eventually, we see that we can write
\be
\textbf{M}^n \ket{i} = \frac{1}{\sqrt{2}} \Big[\cos{\alpha} \, \lambda_{1}^{n} \, \Phi^{+} + \sin{\alpha} \, \lambda_{2}^{n} \, \Psi^{-} + r^{n} \, (a_n \, \Phi^{-} + b_n \, \Psi^{+})\Big],
\label{Mni}
\ee
with $a_n, b_n$ real coefficients that can be easily computed.
After normalization, depending on the eigenvalue that
dominates for $n \rightarrow \infty$ ($\lambda_{1}^{n}$, $\lambda_{2}^{n}$, or $r^n$),
the asymptotic state will be $\Phi^{+}, \Psi^{-}$, or a state of the form 
$\cos{\xi} \, \Phi^{-} + \sin{\xi} \, \Psi^{+}$, with the angle $\xi$ obtained in the limit.
The three states are maximally entangled states, and entanglement saturation is ensured.

To show even more explicitly how the mechanism works, we consider the ultra-relativistic limit
(relevant
from a physical point of view), in which the matrix (\ref{MB}) is Hermitian. In this case, the
eigenvectors are the four Bell states. Again,
for the initial state $\ket{i} = \cos{\alpha} \ket{RR} + \sin{\alpha} \ket{RL}$, one can straightforwardly compute the
normalized state after $n$ iterations, obtaining
\be
\ket{f_n}_{\theta} = \frac{\cos{\alpha} \, [\Phi^{+} + \Phi^{-}]
+ \sin{\alpha} \, [(1 + (\cos{\theta})^2)^n \, \Psi^{+} 
+ (2 \cos{\theta})^n \, \Psi^{-}]}{\sqrt{2 (\cos{\alpha})^2 + (\sin{\alpha})^2 \, (1 + (\cos{\theta})^2)^{2 n}
+ (2 \cos{\theta})^{2 n}}}.
\label{NItBUR}
\ee
From the inequality $1 + (\cos{\theta})^2 \ge 2 \cos{\theta}$, one has that, for $\theta \neq 0, \pi, 2 \pi$, the asymptotic state is $\Psi^{+}$.

We can finally see that the same mechanism works also if we randomly change the angle at each step.
Also in this case the entanglement saturation can be
obtained by the expansion of any state in terms
of the eigenvectors of the scattering matrix, the unique
difference being the fact that the eigenvalues can change by varying the
scattering angle.
Consider ideed the Bhabha scattering in the ultrarelativistic regime,
and suppose that the random distribution of scattering angles after $n$ iterations
is given by $k$ angles $\{\theta_i\}_{i = 1}^{k}$, with $k \le n$,
and where the angle $\theta_i$ is repeated $k_i$ times,
with $\sum_{i = 1}^{k} k_i = n$.
Then, if the initial state is $\ket{RL}$, we obtain at the $n$-th step the normalized state
\be
\ket{f_n}_r=\frac{\Big[\prod_{i = 1}^{k} (1 + (\cos{\theta_i})^2)^{k_i}\Big] \, \Psi^{+} 
+ \Big[\prod_{i = 1}^{k} (2 \cos{\theta_i})^{k_i}\Big] \, \Psi^{-}}{\sqrt{\Big[\prod_{i = 1}^{k} \, ((1 + (\cos{\theta_i})^2)^{k_i}\Big]^2 +
\Big[\prod_{i = 1}^{k} (2 \cos{\theta_i})^{k_i}\Big]^2}}.
\ee
Since, again, $1 + (\cos{\theta_i})^2 \ge 2 \cos{\theta_i} \, \forall i$,
for $n \rightarrow \infty$ the coefficient of $\Psi^{+}$ dominates, and $\Psi^{+}$
is the asymptotic state. 

As last remark, we see that the asymptotic saturation of entanglement is
implied only by the form of the scattering matrix, but the identification of the dominant
eigenvalue requires the explicit expressions of the scattering amplitudes.

\section{Processes involving photons}

If photons are involved in the scattering process, the complete conservation
of maximum entanglement is lost \cite{BlasoneSE2, NoiChaos}. Analogously, the asymptotic
saturation of the entanglement is
reduced, or canceled, due to the presence of two different statistics.
In fact, if we consider the scattering process $e^{-} e^{+} \rightarrow \gamma \gamma$, 
assuming as initial state $\ket{RL}$, in the ultra-relativistic regime one obtains 
the asymptotic state $\Psi^{-}$,
while from the initial state $\ket{RR}$ a not maximally entangled state is generated.
This result can be traced back to the fact that not all the eigenvectors of the matrix describing this process are Bell states or the suitable combination
of Bell states found in the case of 2 fermions $\rightarrow$ 2 fermions scattering. 
Another peculiarity of this process  is that the second property, 
self-similarity by raising to a generic power,
is not fully satisfied.
Finally, in the case of the Compton process,
no eigenvector of the corresponding scattering matrix
is a Bell state or a suitable combination of Bell states, and entanglement saturation is never obtained.

\section{Conclusions}

Dynamics of entanglement after a sharp filtering at fixed momentum in QED scattering processes
is realized by a positive operator valued measurement on the final state after scattering. Although this procedure does not correspond to a unital quantum channel, 
for which a quantum version of the H-Theorem holds, if 2 fermions $\rightarrow$ 2 fermions scattering is considered, maximum entanglement is always preserved. 
The action of  scattering processes is realized in this case by quantum maps,  expressed in terms of scattering amplitudes. 
Iteration of the maps works
in such a way as to realize maximal inner symmetry: for example, when starting from a separable state, the iteration process forces coefficients
in the computational basis to become equal, leading to the loss of information
about the inner structure of the state. The increase of entanglement during the iterations is not necessarily monotonic because, in
our framework, also entanglophobous states can be met, but eventually entanglement always saturates. This result is due to properties fulfilled by the maps: in fact, they do not only ensure the conservation of maximum entanglement, but also their 
iteration selects as asymptotic state a maximally entangled state.
The validity of the above property is
reduced, more or less extensively depending on the process, when not only fermions
but also photons are involved in the scattering interaction.

A remarkable aspect is the fact that the general forms of the matrices defining the maps, independently of the
explicit expressions of their elements, are sufficient to identify the invariant sets, including the set of the maximally
entangled states, and to ensure the characteristics of the whole entanglement dynamics, including the entanglement's asymptotic behavior. Since the forms of the matrices are due
to the relations among the scattering amplitudes in each process, in turn determined by the characteristics of the
related fundamental interaction, future studies will be devoted to identify underlying symmetries which can be the
ultimate origin of the entanglement’s behavior \cite{Kowalska,McGinnis:2025brt}, and to investigate the possible existence of similar phenomena in weak and strong
interactions.


\end{document}